\documentclass[12pt,dvips]{article}
\usepackage{epsfig}

\title{\vspace{-5.0cm}
\begin{flushright}
{\normalsize UNIGRAZ-}\\
\vspace{-0.3cm}
{\normalsize UTP-}\\
\vspace{-0.3cm}
{\normalsize 18-07-96}\\
\end{flushright}
\vspace*{2.5cm}
U(1) Gauge Theory with Villain Action on Spherical Lattices}
\author{\bf C. B. Lang\protect\footnotemark[1]~ 
and P. Petreczky\protect\footnotemark[2]\\~\\
Institut f{\"u}r Theoretische Physik,\\
Universit{\"a}t Graz, A-8010 Graz, AUSTRIA}
\date{\today}
\begin{document}
\thispagestyle{empty}
\maketitle
\begin{abstract}
We have studied the U(1) gauge field theory with Villain (periodic
Gaussian) action on spherelike lattices. The effective size of the
systems studied ranges from 6 to 16. We do not observe any 2-state
signal in the distribution function of the plaquette expectation value
at the deconfining phase transition.  The observed finite-size scaling
behavior is consistent with a second order phase transition. The obtained
value of the critical exponent is $\nu =0.366(12)$ and thus neither
Gaussian ($\nu = 0.5$) nor discontinuous ($\nu=0.25$) type, indicating
a nontrivial continuum limit.  
\end{abstract}

\vspace*{0.5cm}

\renewcommand{\thefootnote}{\fnsymbol{footnote}} 
\footnotetext[1]{e-mail: cbl@kfunigraz.ac.at}
\footnotetext[2]{on leave from L. Kossuth University, Debrecen, Hungary}

\newpage
\section{Introduction}

In this work we study the Villain formulation \cite{Vi75} of the U(1)
pure gauge theory on 4D lattices with spherelike topology.  We were
motivated by recent results obtained in the theory with Wilson action
(with and without a charge two coupling $\gamma$) on spherelike
lattices \cite{LaNe94,JeLaNe,JeLaNe96ab}.  It turned out that on such
lattices there are no 2-state signals even on the Wilson line
($\gamma=0$) \cite{JeLaNe}.  For hypercubic lattices with periodic
boundary conditions (the usual torus geometry) one finds such signals
\cite{JeNeZe83,EvJeNe85,KeReWe,BoLiSc} even at sizable negative values
of $\gamma$ \cite{EvJeNe85}; under certain assumption a tricritical
point (TCP) was predicted at $\gamma =-0.11$ \cite{EvJeNe85}.  It has
been argued \cite{LaNe94} that the disturbing effects may be due to the
interplay of periodic boundary condition with the important r\^ole of
the topological monopole excitations in the transition. This issue is
still unsettled, however.  In case the two-state signal persists in an
infinite-volume limit, the transition is first order; if the signal is
spurious, it prevents a careful finite-size scaling (FSS) analysis of
the critical transition.  Up to date taking a continuum limit has not 
been possible due to this problem.

We hope that in this situation simulations with the Villain action help
to clarify some of the problems mentioned above. This formulation  is
of certain interest because it lends itself  to theoretical analysis.
In this form of the theory the partition function can be decomposed
into Gaussian fluctuations and monopole excitations \cite{BaMyKo77}.
There are further relationships to the noncompact U(1) Higgs model in
the limit of large negative squared bare mass (frozen 4D
superconductor) \cite{Pe78,FrMa} and to an effective string theory
equivalent to that model \cite{PoStPoWiZu,PoWi91}.  The action obeys
reflection positivity.

The leading terms of a character expansion of the Villain action shows
that the  phase transition (PT) is near $\beta_W=1.16$ 
(we denote by $\beta_W$ the coupling in the Wilson action) and
$\gamma=-0.22$.  Although this value of $\gamma$ is below the value for
a conjectured TCP \cite{EvJeNe85} metastability signals were observed
(for torus b.c.) here, too \cite{LaRe87,GrJaJe85}. The PT lies in a
hypersurface of PTs for the other actions mentioned.  It is therefore
of interest, whether the two-state signals vanish for spherelike
lattice for the Villain action as well, and, if this is the case,
whether the critical behavior is in the same universality class.

We proceed in the following way: In section 2 we discuss the details
and implementation of the Villain action, in section 3 the lattices
with spherelike topology are introduced and the general strategy of the
simulation and the FSS analysis are presented.
Section 4 summarizes our conclusions.

\section{The Villain action}

The Villain or heat-kernel action is defined through
the Boltzmann factor per plaquette 
\begin{equation}
\exp{\left(-S_P(\beta,U_P)\right)} \equiv K(1/\beta,U_P),
\end{equation}
and is the solution of the diffusion equation in group space
\begin{equation}
\Delta K(t,U)=-{d \over dt}K(t,U),
\end{equation}
with the boundary conditions
\begin{equation}
K(0,U)=\delta({\bf 1},U),\;K(\infty,U)=1.
\end{equation}
It connects the constant distribution at strong coupling ($t=\infty$)
with a distribution peaked at the group unit element in the weak coupling
limit ($t=0$).

The heat equation is implemented by introducing the metric tensor
and the Laplace-Beltrami operator on the group manifold.
In the case of the U(1) gauge group the solution can be written as
\begin{equation}
\exp(-S_P)=\sum_{n=-\infty}^{\infty}\exp \left[ -{\beta \over 2} 
{\left(\theta_P-2 n \pi \right)}^2 \right]
\end{equation}
where $\theta_P$ is the angle of the plaquette variable 
$U_P=\cos \theta_P$ and given as a sum over
all link angles $\theta_{x,\mu} \in [-\pi,\pi]$
\begin{equation}
\theta_P \equiv \theta_{x,\mu \nu}=\theta_{x,\mu}+\theta_{x+\mu,\nu}-
\theta_{x+\nu,\mu}-\theta_{x,\nu}
\end{equation}
In the simulation $S_P$ has been calculated by interpolating pretabulated
values.

The total action is defined through
\begin{equation}
\exp{\left( -S(\beta)\right)} = \prod_P \exp{\left( -S_P(\beta,U_P)\right)}
\end{equation}
and the observables (quantum averages) are given 
as expectation values with regard to the Gibbs measure $\prod_{x,\mu}
dU_{x,\mu} \exp{\left( -S(\beta)\right)}$.
Useful observables are the internal energy
\begin{equation}
E_{HK}=-\langle\sum_P {\partial S_P \over \partial \beta}\rangle
\end{equation}
and the specific-heat
\begin{equation}
C_{HK} = {\partial E_{HK} \over \partial \beta }=
\langle {\left(\sum_P {\partial S_P \over \partial 
\beta}\right)}^2 \rangle
-\langle\sum_P{\partial S_P \over \partial \beta}\rangle^2
-\langle\sum_P{\partial^2 S_P \over \partial \beta^2}\rangle .
\end{equation}
This form differs from the usual definition due to the
non-linear dependence of $S_P$ on $\beta$. Note, that $\beta$ is
different from the coupling $\beta_W$ used in the Wilson-action. 

In our calculations we measure the expectation value of the Wilson
plaquette variable
\begin{equation}
E\equiv \langle\sum_P w_P\,\mbox{Re}\, U_P\rangle
\end{equation}
and its distributions (the weight factors are due to correction factor
for the spherical lattice shape and are discussed below).  For the
investigation of the PT both sets of observables are equally suited; we
find the later set more useful, however.  The  histogram analysis along
the lines of Ferrenberg and Swendsen \cite{FeSw} is implemented in that
variable $E$ and the analytic continuation therefore is in the
conjugate coupling variable $\beta_W$.  From the histograms in $E$
corresponding higher order cumulants are determined,
\begin{eqnarray} \label{cumdefs}
c_V(\beta,L)&=& ~~\frac{1}{6V} \langle (E-\langle E\rangle )^2
\rangle  ,\\
V_{CLB}(\beta,L)&=& -\frac{1}{3} \frac{\langle (E^2-\langle
E^2\rangle )^2\rangle}{\langle E^2\rangle^2}  ,\\
U_4(\beta,L)&=& \frac{\langle (E-\langle E\rangle)^4\rangle}
{\langle(E-\langle E\rangle )^2\rangle^2}   .
\end{eqnarray}
(For simplicity we call $c_V$ specific-heat, too.)
The positions and values of their respective extrema are used for the
FSS analysis. 

\section{Numerical simulation and  results}

We have performed simulation on the following lattices with spherelike
topology:  
\begin{itemize} 
\item SH[N], the surface of 5D hypercubes of size $N^5$;
\item S[N], with the topology of SH[N] but introducing
weight factors for the plaquette action:  $S = \sum_P w_P S_P $,
correcting for spherical shape. i.e. distributing the curvature over
the lattice (details about the geometry and these factors are given
in \cite{JeLaNe96ab}).  
\end{itemize}

Most of our simulations were done on S[N]  lattices with $N=$4, 6, 8,
10; these lattices have roughly the same number of variables as $6^4$,
$9^4$, $12^4$, $16^4$ hypercubic lattices.  In average we have
performed about half a million sweeps for each lattice size, using a
3-hit Metropolis algorithm.  We have measured the values and
distribution functions (histograms) for $E_{HK}$ and $E$ and found no
indication of double peaks.  Unfortunately a multihistogram analysis
combining various histograms determined at different values of $\beta$
is not applicable here because of the specific form of the action:
$S_P$ does not depend linearly on $\beta$.  However, we are able to do
the analysis based on single histograms in $E$ in the vicinity of the PT
and extrapolate into the ``Wilson direction'' $\beta_W$ based on the
individual histogram,
\begin{equation}
\langle f(E) \rangle = \frac{1}{A}\sum_E h(E,\beta) f(E) e^{-\beta_W E},
\; A\equiv \sum_E h(E,\beta) e^{-\beta_W E} ,
\end{equation}
where $h(E,\beta)$ denotes the histogram entry at $E$. The number of
bins was taken sufficiently large (typically $>2000$) to 
exclude systematic errors.

For each lattice size we try to simulate as close as possible to the
peak position of the specific-heat.  For this aim we first determined
the cumulants as functions of $\beta_W$ from histograms at various
values of $\beta$.  This preparatory analysis was done at the
statistics of 100K sweeps.  From this we infer a relationship between
$\beta_W(peak)$ and $\beta$ and determine the presumed value of the
corresponding pseudocritical coupling $\beta_0$.  These are given in
Table 1.  Then we performed long runs (about 500K sweeps) at these
couplings. The final results are histograms determined almost on top of
the specific-heat peak. The individual histogram analysis allows us to
extrapolate away from this point in coupling space into the direction
$\beta_W$ and the table also shows this distance to the actual peak
position of the specific-heat.  This provides further (small)
corrections.

For our analysis we use the values of the cumulants at their
(extrapolated) maxima or minima, which agree within the errors with the
values measured directly at $\beta_0$.

\begin{table}
\begin{center}
\begin{tabular}{rrr}
\hline
S[N]	&$\beta_0$ &$\Delta \beta_W$\\
\hline
\hline
S[4]	&0.63687&   -0.00029(46)\\
S[6]	&0.64550&    0.00060(28)\\
S[8]	&0.64766&    0.00067(10)\\
S[10]	&0.64867&    0.00017(14)\\
\hline
\end{tabular}
\end{center}
\caption{Couplings $\beta_0$ where the long runs were performed;
we also give the extrapolation distance to the peak position
of the specific-heat in direction of $\beta_W$ (as determined a 
posteriori from the histogram analysis).}
\end{table}

\begin{figure}[htf]
\begin{center}
\epsfig{file=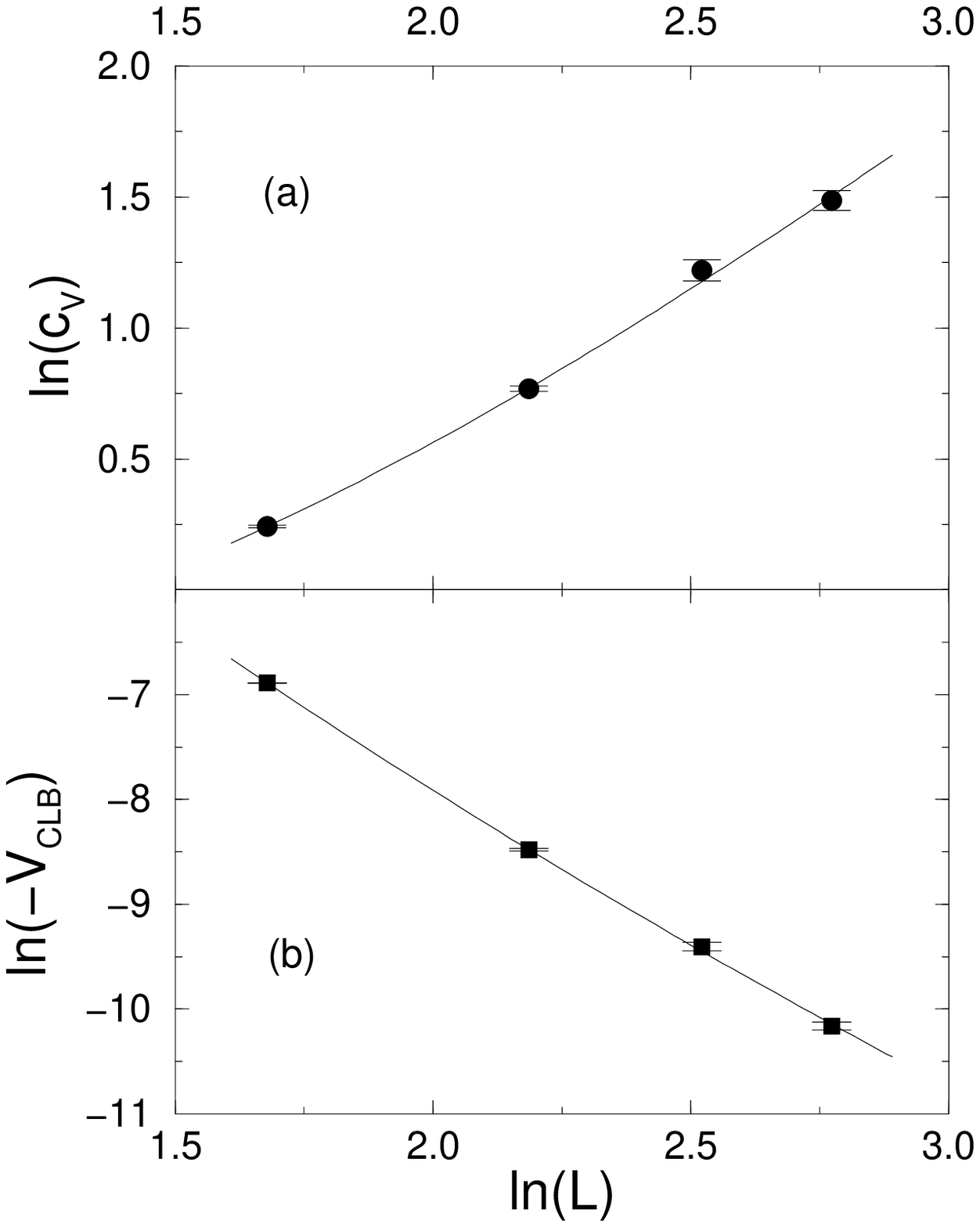,width=9truecm}
\end{center}
\caption{The logarithm of the extrema values of (a) the specific-heat
and (b) the Challa-Landau-Binder $V_{CLB}$ cumulant vs $\ln L$; the
fits represent the leading FSS behavior given in (\ref{FSScum}).} 
\end{figure}

The final histograms are similar to those determined for the Wilson
action \cite{JeLaNe96ab} and show no 2-state signal. (Note that these
are individual histograms and not the result of a combination.) We
proceed to study the FSS of the extrema value of the specific-heat
(maximum) and  Binder cumulants $V_{CLB}$ and $U_4$ (minima, cf.
\cite{LaNe94,JeLaNe} for more details on the definitions and FSS
properties) to check for consistent critical (i.e. second order ) FSS
behavior, which in leading order may be parameterized
\begin{eqnarray}\label{FSScum}
c_{V,max}(L)   &\simeq& a_c + b_c L^{\alpha/\nu} ,\\
V_{CLB,min}(L) &\simeq& (a_v + b_v L^{\alpha/\nu})L^{-4} ,\; \\
U_{4,min}(L)   &\simeq& a_u + b_u L^{-\alpha/\nu} .
\end{eqnarray}
In our analysis we assume validity of the hyperscaling relation 
$\alpha=2-D \nu$ ($D=4$ in our case). The length scale 
$L\equiv V^\frac{1}{4}$, where the
volume $V=\frac{1}{6}\sum_P w_P$, roughly proportional to the number of sites
(in fact, this definition gives the number of sites in the situation of the usual
hypercubic lattice with torus boundary conditions).

A log-log plot of the peak value of the specific-heat as a function of
L is shown in Fig.  1a.  From the fit we obtain the value ${\alpha
/\nu}=1.479(135)$ and $\nu=0.365(9)$ ($\chi^2/d.f.=1.3$), which is far
from what one expects in the case of a first order PT, where ${\alpha/
\nu}=4$ and $\nu=1/D=0.25$.  This value is also compatible with the
values obtained in \cite{JeLaNe,JeLaNe96ab}.  The cumulant
$V_{CLB,min}$ is shown in Fig 1b; its FSS behavior is in good agreement
with that of the specific-heat.  The fit due to (\ref{FSScum}) leads to
$\nu=0.359(10)$ ($\chi^2/d.f.=0.8$). The cumulant $U_{4,min}$ is
compatible with the expected FSS; it approaches a constant $\simeq
2.75(3)$ but has too large errors to determine the nonleading scaling
term, although it is consistent with the values of $\nu$ determined
from the other quantities.

Let us discuss briefly the FSS of the pseudocritical couplings (as
summarized in Table 1).  Generally the FSS of pseudocritical couplings
may be written in the form
\begin{equation}
\beta(V)=\beta(\infty)+cL^{-\lambda},
\label{betc}
\end{equation}
where $\lambda$ is the so-called shift-exponent. For many models
$\lambda=1/\nu$, but this is not necessarily so in general (see the
discussion in \cite{Ba83}; a recent study of the 2D Ising model on
lattices with spherelike topology yielded $\lambda \neq 1/\nu$
\cite{HoLa96}). Therefore the FSS behavior of the pseudocritical coupling
is not necessarily suitable for an independent determination of $\nu$.
A fit to the values in Table 1 gives $\lambda= 2.71(26)$ -- consistent
with  $\lambda=1/\nu$ within the large error bars -- and
$\beta(\infty)=0.6496(3)$.  Note, that due to the geometry correction
factors this latter value does not have to agree with the corresponding
value for torus topology.  A more detailed discussion in a similar
context can be found in \cite{JeLaNe96ab}.

\begin{figure}[htf]
\begin{center}
\epsfig{file=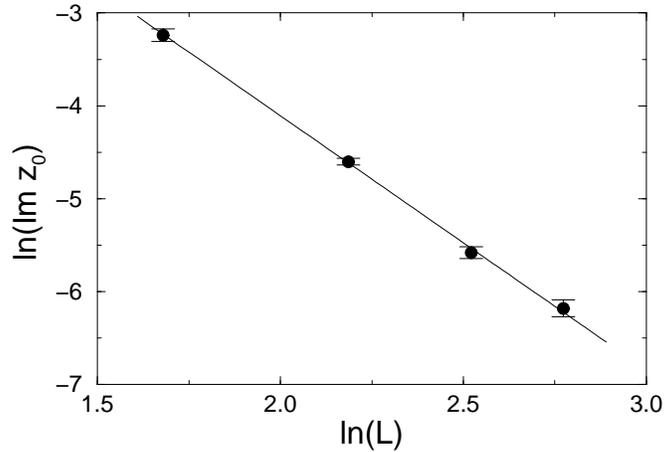,width=9truecm}
\end{center}
\caption{The imaginary part of the closest Fisher zero as the function
of the size $L$.}
\end{figure}
 
Another efficient way to determine the  exponent $\nu$ lies in
the FSS of the imaginary part of the (Fisher) zeroes $z_0$ of the
partition function (in the complex coupling plane) closest to the real
axis. The corresponding plot is shown on Fig.2; the expected scaling is
$\mbox{Im} z_0 \propto L^{-1/ \nu}$.  From the fit one obtains
$\nu=0.366(12)$, using all lattice sizes ($\chi^2/d.f.=0.5$).  This is
in excellent agreement with the values obtained from the cumulants.

We have also performed simulations on SH[N] lattices with $L=4, 6, 8$.
Since here the statistics was only 20\% of that for the S[N] lattices
we do not quote errors and did not analyze these data in more detail.
The main result is that also for SH[N] we do not see any 2-state
signal. The pseudocritical couplings $\beta_0$ are 0.64, 0.64375, 0.645
for corresponding lattice sizes.  The peak values of the specific-heat
are 1.48, 2.77, 3.32 correspondingly. All these results are close to
those obtained on S[N] lattices.

\section{Conclusion}

We have studied the compact U(1) gauge theory with the Villain action
on 4D lattices with spherelike topology.  We do not observe any
metastability (i.e. 2-state signal) in the distribution function of the
internal energy. The FSS for various quantities is consistent with
critical behavior. The obtained value of the critical exponent  $\nu$
is compatible with neither the Gaussian value 0.5 nor the
``discontinuity'' value 0.25.  In our determination for the Villain
action we get values in the range 0.36--0.37, a smaller range of
values, but in perfect agreement with those obtained in
\cite{JeLaNe96ab} with mixed action on spherical lattices. Note that
these actions are at quite separate points in the space of plaquette
actions given by a character expansion.

These results indicate that below the Wilson line the deconfining PT is
of second order and both the mixed action and the Villain action
belong to one universality class of models with a nontrivial
continuum limit.

\subsection*{Acknowledgment}
C.B.L. thanks J. Jers{\'a}k and T. Neuhaus for discussions.
P.P. was supported by a TEMPUS grant under the EMSPS exchange
program. The computations have been performed  on the Parallel
Compute Server of KFU of Graz.

\end{document}